# Coexistence of Van Hove Singularities and Pseudomagnetic Fields in Modulated Graphene Bilayer

Jana Vejpravova[1*], Barbara Pacakova[2,3], Mildred S. Dresselhaus[4 1#], Jing Kong[4], and Martin Kalbac[3**]

[1] Department of Condensed Matter Physics, Faculty of Mathematics and Physics, V Ke Karlovu 5, 121 16 Prague 2, Czech Republic
[2] Norwegian University of Science and Technology, Department of Physics, Gamle fysikk, E3, R-149, Gløshaugen, Norway
[3] J. Heyrovsky Institute of Physical Chemistry of the Czech Academy of Sciences, Department of Low-Dimensional Materials, Dolejskova 3, 18200 Prague 8, Czech Republic
[4] Department of Electrical Engineering and Computer Science, Massachusetts Institute of Technology, Cambridge, Massachusetts 02139, USA

E-mail: *jana@mag.mff.cuni.cz; **martin.kalbac@jh-inst.cas.cz

## Abstract

The stacking and bending of graphene are trivial but extremely powerful agents of control over graphene's manifold physics. By changing the twist angle, one can drive the system over a plethora of exotic states via strong electron correlation, thanks to the moiré superlattice potentials, while the periodic or triaxial strains induce discretization of the band structure into Landau levels without the need for an external magnetic field. We fabricated a hybrid system comprising both the stacking and bending tuning knobs. We have grown the graphene monolayers by chemical vapor deposition, using $^{12}C$ and $^{13}C$ precursors, which enabled us to individually address the layers through Raman spectroscopy mapping. We achieved the long-range spatial modulation by sculpturing the top layer ($^{13}C$) over uniform magnetic nanoparticles (NPs) deposited on the bottom layer ($^{12}C$). An atomic force microscopy study revealed that the top layer tends to relax into pyramidal corrugations with $C_3$ axial symmetry at the position of the NPs, which have been widely reported as a source of large pseudomagnetic fields (PMFs) in graphene monolayers. The modulated graphene bilayer (MGBL) also contains a few micrometer large domains, with the twist angle ~ 10 °, which were identified via extreme enhancement of the Raman intensity of the G-mode due to formation of Van Hove singularities (VHSs). We thereby conclude that the twist induced VHSs coexist with the PMFs generated in the strained pyramidal objects without mutual disturbance. The graphene bilayer modulated with magnetic NPs is a non-trivial hybrid system that accommodates features of twist induced VHSs and PMFs in environs of giant classical spins.

---

[1]   # In memory of the deceased professor M.S. Dresselhaus, who passed away unexpectedly in 2017, we decided to release the study in light of current interest in extraordinary physics of twisted graphene bilayers. She collaborated on the work since summer 2012 and significantly contributed via her interpretation of the Raman spectra, making many inspiring comments on the very first version on the manuscript, including the importance of the twist angle in the bilayer.

## 1. Introduction

Exotic condensed matter physics, represented by strong electron correlations (SEC), large spin–orbit coupling, and complex topology of electronic states is usually a privilege of systems with heavy electrons [1]. However, many of the unconventional states originated by the SEC, including the Mott insulating regime and unconventional superconductivity, can be stabilized in a simple material composed of a single element—a graphene bilayer [2, 3]. As demonstrated many times [4–7], the twist or shear in the bilayer has a significant influence on the interlayer hopping and thus on the renormalization of the band structure. In the low angle limit ($< 3°$), the twist-induced Van Hove singularities (VHSs) dominate the band structure, while in the large twist angle limit ($> 20°$), the individual graphene monolayers (1-LG) behave as if they are independent [6, 8, 9]. The twist between the two 1-LG layers is also responsible for the change of the low-energy pseudospin texture, which is accompanied by the change of the quasiparticle chiralities [10]. Thanks to the misoriented stacking, the creation of in-plane pseudo-magnetic fields (PMFs) exceeding $10^3$ T was predicted [10–12]. Recently, a PMF with variable spatial distribution and intensity has been reported for the 1-LG on black phosphorus [13].

In addition to the proper twist providing appropriate moiré potentials, the controlled deviations from flatness, such as the periodic or triaxial strains acting on the graphene lattice without lowering its symmetry [14, 15], also cause discretization of the band structure. This gives rise to the pseudo-Landau levels due to the PMFs [14–17]. A recent theoretical study also demonstrated that a rippled 1-LG exhibits a superposition of local ferromagnetic and antiferromagnetic ordered states with a zero global ferromagnetic order parameter [18]. The formation of a gapped quantum Hall local pseudo-ferromagnetic state with positive and negative spin polarization in regions with positive and negative effective gauge fields, respectively, was also predicted [18].

The most common topographic corrugations randomly or intentionally generated in graphene include the ripples, the wrinkles, and the collapsed forms of the wrinkles, such as folds and crumples (for a brief introduction to the topic, please see the review by Deng and Berry [19] and the references therein). All of these introduce non-trivial strain to the graphene lattice associated with the above discussed phenomena [14–18].

A straightforward approach toward tailoring the 1-LG topography is the application of tiny objects such as nanoparticles (NPs) or nanopillars as sources of corrugations [20–23]. Their geometry on the substrate determines the final topography of the 1-LG and thus promotes the coveted band structure modifications with very high spatial precision.

The open question is whether both tuning knobs—the stacking and the strain—can be applied simultaneously to reach a system with multiple exotic features coexisting in the same nanostructure.

In our work here, we built up modulated graphene bilayer (MGBL) to explore this possibility. We used uniform magnetic NPs deposited on the bottom layer in a proper geometry to achieve local corrugation of the top layer, giving rise to the PMFs. The topographic variation results in both a spatially distributed doping (position of the Fermi level) and a spatially distributed strain in both 1-LG layers, which we addressed through Raman spectroscopy. Using state-of-the-art isotope labeling, we were able to address both layers individually. In many areas of the MGBL, we achieved a very low twist angle between the layers, giving rise to the formation of VHSs, which we clearly identified via Raman spectroscopy. We proved that the VHSs and the PMF could coexist in the same area of MGBL, which is defined by the size of the laser spot.

## 2. Experimental

The sample preparation concerns the synthesis of uniform magnetic NPs through a hydrothermal method and the isotopically labeled graphene through chemical vapor deposition (CVD) [24].

The NPs with a mean diameter of the oxide core observed by the transmission electron microscopy (diameter ~ 9 nm), coated with covalently bonded oleic acid (thickness ~1 nm), were prepared in analogy to those described in previous work [25]. In brief, the following items were added to a 100 ml beaker that was then stirred: NaOH (14 mmol), water (4 ml), pentanol (5 ml), and oleic acid (18 mmol). From this mixture, a clear solution was obtained. Then, a water solution of $Fe(NO_3)_3 \cdot 9H_2O$ (2 mmol) and $FeCl_2 \cdot 4H_2O$ (4 mmol) was added, and the total amount of water was 15 ml. The mixture was then stirred under sonication for 3 min and subsequently left for 15 min to separate the colorless water phase and the dark brown organic phase. The water phase was discarded and the organic phase (containing 5 ml of pentanol and iron oleates) was moved into a 50-ml Teflon tube, sealed in an autoclave under Ar atmosphere and kept at 180 °C for 10 h. The particles thus obtained were redispersed in hexane (10 ml) and precipitated several times with ethanol (10 ml). The final precipitate was isolated by a magnet and dispersed in hexane; large aggregates were removed through centrifugation.

In the CVD process, the Cu foil was heated to 1000 °C and annealed for 20 min under flowing $H_2$ (50 sccm). The foil was then exposed to $^{12}CH_4$ or $^{13}CH_4$ for 20 min, leaving the hydrogen gas on at the same flow rate. The etching of the top layers was realized by switching off the methane and leaving on the hydrogen gas for an additional 1–20 min at 1000 °C. Finally, the substrate was quickly cooled down under $H_2$. The MGBL was obtained through the following steps (a schematic representation of the process is shown in Figure 1). First, the CVD-grown $^{12}C$-1-LG was transferred onto the $SiO_2$/Si substrate using polymethylmethacrylate (PMMA) according to a reproducible procedure that was previously reported [26]. The sample was annealed in oxygen at 300 ºC for 1 h. The $^{12}C$-1-LG was then decorated with the NPs through the spin coating of their dispersion in hexane (concentration of 0.01 mol Fe/ml). The concentration of NPs on the $^{12}C$-1-LG was evaluated using atomic force microscopy (AFM). Finally, the CVD-grown $^{13}C$-1-LG was transferred on top of the $^{12}C$-1-LG decorated with the NPs using the same procedure as for the $^{12}C$-1-LG transfer. The top layer was transferred only onto part of the bottom layer to obtain reference areas for the sample for complex evaluation of the doping, strain, and topographic corrugations.

The samples were then characterized using AFM and Raman micro-spectroscopy. The AFM images (size of 25 $\mu m^2$) were captured under ambient conditions in the standard tapping mode with a Veeco Multimode V microscope equipped with a JV scanner with a resolution of 1024 lines and a scan rate of 0.8 Hz. A fresh high-quality etched silicon probe (k = 3 N/m, $f_0$ = 75 kHz, nominal tip radius = 8 nm) created by Bruker was used.

The AFM data were further subjected to advanced analysis with the help of Gwyddion software. The values of the root-mean square roughness ($\sigma_p$) of the bare substrates, decorated substrates, and the smooth and delaminated parts of the 1-LG were determined. The value of the relative areas attributed to the delaminated 1-LG (*AW*) were finally extracted for each image. A complete description of the procedure can be found in the recent work of Pacakova et al. [27] and in the Supplementary Information file, section 1.

The Raman spectra were acquired by a LabRam HR spectrometer (Horiba Jobin-Yvon) using 633 nm He/Ne excitation. The spectral resolution was around 1 $cm^{-1}$. The spectrometer was interfaced to a microscope (Olympus, 100× objectives) so the spot size was around 1$\mu m^2$. The typical laser power measured at the sample was around 1 mW. The Raman maps (size of 30×30 $\mu m^2$) were recorded with steps of 1 $\mu m$. All bands were fitted either as single pseudo-Voigt functions, $\Omega(I,\omega)$, or their superposition; for an explicit form of the relation, please see eq. S1 in the Supplementary Information file. The Raman maps (typically containing 900 spectra) were analyzed using a homemade routine in Octave, enabling automated fitting in batches and the setting of relevant constraints between the parameters [20].

In order to have a simple comparison of the Raman data, the Raman spectra obtained on the $^{13}C$-labelled 1-LG were recalculated as expressed, using equation (1):

$$\frac{\omega_{12}-\omega_{13}}{\omega_{12}} = 1 - \sqrt{\frac{12+c^{13}}{12+p}}, \qquad (1)$$

where $\omega_{12}$ and $\omega_{13}$ are the Raman shifts of the natural $^{12}C$ and $^{13}C$ enriched samples, respectively. $c^{13}$ is the natural abundance of $^{13}C$ in the graphene grown with the $^{12}CH_4$ precursor, which is equal to 0.0107. $p$ is the purity of the $^{13}C$ in the enriched sample, determined by the isotope purity of the $^{13}CH_4$ precursor (here, $p$ = 0.99).

## 3. Results and discussion

*3.1 Topography analysis*

First, we investigated the topography of the hybrid using AFM. We analyzed the sample before and after the deposition of the $^{13}C$-1-LG on the $^{12}C$-1-LG decorated with NPs (the mean diameter of the NPs deposited on a reference flat substrate is 9.7 nm; see Figure S1). The initial concentration of the NPs in the dispersion (about 0.01 mol Fe/ml) was adjusted according to a previous study [16, 20] in order to obtain an optimal concentration of the NPs on the substrate of around 100–150 NPs/$\mu m^2$, thereby enabling a controlled contact regime for the top layer ($^{13}C$-1-LG) after the second transfer. The prepared MGBL samples were carefully checked by analysis of the AFM images as described previously [20, 27]. The optimized configuration of the NPs on the substrate reveals the mean distance of around 50 nm between the NPs, which is supposed to ensure the formation of isolated topographic features around the NPs and have a negligible effect on their mutual dipole–dipole interactions [28].

The typical AFM images of the resulting MGBL structure are shown in Figure 2. Panel (a) represents a large-scale image, which reveals two dominant landscapes—homogeneously distributed pyramidal corrugations at the position of the NP and flat areas. Panel (b) depicts a detailed image of such features; the maximum height and the base edge size are each about 11 nm. Most of the isolated objects conform to the $C_{3v}$ symmetry; however, some complex branching structures with comparable height can be also detected. In addition, a net of tiny wrinkles with a height below 5 nm and a length of a few dozen nanometers contribute to the final landscape of the MGBL.

We further compared the three different representative areas of the sample: $^{12}$C-1-LG decorated with the NPs, $^{13}$C-1-LG transferred over the NPs on the substrate, and 2-LG. The corresponding AFM images and height histograms are shown in Figures S2 and S3, respectively. The 1-LG that was transferred over the decorated substrate shows AFM patterns that were previously observed on monolayer graphene that had been deposited on decorated substrates [16, 18]. The deviation from the flatness is represented by the relative delaminated area, *AW*, obtained through grain analysis of the AFM data [16, 22]. The *AW* value was found to be the lowest for the $^{12}$C-1-LG (*AW* = 16%); *AW* increases moderately for $^{13}$C-1-LG (*AW* = 19%) and reaches about 29% for the 2-LG area. The values of *AW* derived from $^{13}$C-1-LG (~ 120 NPs/$\mu m^2$) correspond very well with the 1-LG transferred onto a SiO$_2$(300 nm)/Si substrate decorated with NPs of similar size and density [20]. We note that one can find wrinkles on the samples even without adding NPs on substrate, however the number and character of wrinkles induced by NPs is significantly different [20, 22, 27].

Based on our previous results [20, 27], the typical level of delamination in the CVD-grown 1-LG transferred onto the SiO$_2$/Si substrate is below 5%, and the main delamination is located in the few micrometer-long wrinkles (half carbon nanotube-like), which likely developed from the transfer process. In the MGBL structure, the reference bottom layer ($^{12}$C-1-LG) shows a higher delamination level (*AW* = 16%). This effect can be attributed to the processing conditions during the second transfer step, in which the PMMA-supported $^{13}$C-1-LG is transferred on top of the NP-decorated $^{12}$C-1-LG and the sample undergoes a second annealing cycle in order to remove the PMMA (for details, please see the Experimental).

*3.2 Raman spectra analysis*

The sample was further subjected to a detailed analysis through Raman mapping. The two reference areas and the principal area of the MGBL structure were examined. The typical Raman spectra in the vicinity of the principal Raman-active modes of the graphene (G and 2D [29]) are shown in Figure 3. For consistency with previous work on the wrinkled and crumpled graphene [20], the G-mode shows a clear fine structure, which can be sufficiently well described using three subbands: $G_1$, $G_2$ and D'. The $G_1$ feature is attributed to the fraction of 1-LG that is in contact with the underlayer, while the $G_2$ peak corresponds to the delaminated fraction. The D' peak comes from defect-assisted scattering [29]. In a similar manner, the 2D band can be decomposed into the adhered and delaminated fractions ($2D_1$, $2D_2$) up to a delamination of about 50% [20]. As the inner structure of the 2D mode is expected to be rather complex due to the local distribution of strain [30], the profile of each of the two subbands was approximated using a single pseudo-Voigt function; the admixture of the Gaussian component is assumed to be a measure of the Raman shift distribution within the probed region (given by the laser spot size of about 1µm). The details on spectral decomposition involve given and representative fits of the Raman spectra, including the integrated area under the D-mode, are given in the Supplementary Information file in Figures S4 and S5, respectively.

Maps of the absolute Raman intensity and Raman shift of the $G_1$ and $G_2$ modes in the 2-LG area of the MGBL structure are shown in Figures 4 and 5, respectively. The analog Raman intensity maps for the reference areas ($^{12}$C-1-LG and $^{13}$C-1-LG) are given in the Supplementary Information file (Figure S6).

The most striking feature is the presence of regions with an extreme enhancement of the $G_1$ mode and a simultaneous drop in the Raman shift of the $G_1$ mode; two examples are emphasized with hexagons in panels (a) and (b) of both Figures 4 and 5, and typical spectra in and out of the VHS regions are represented in Figure 3 with red and blue circles, respectively.

We attribute the observed effect to the low twist angle between the two 1-LG layers, where the giant enhancement of the G-mode Raman intensity is associated with the formation of the VHSs in the twist angle regime of a 1-LG bilayer of about 10º [31, 32]. Regarding the AFM analysis, the mean distance between the NPs and thus between the isolated corrugations is about 50 nm, which essentially represents the lower bound for coherence of the VHSs in the corrugated graphene bilayer.

However, this enhancement is not present in the complementary area of the map for the $G_2$ mode. The explanation for this effect is that since the $G_2$ feature is related to the corrugated fraction of the 1-LG layers, the close-to-ideal band structure is not present, as in the case of the flat (adhered) fraction of the 1-LG. Consequently, the VHSs are not formed as expected. In fact, this observation simply underlines the VHSs–PMFs coexistence scenario.

Finally, the maps of the sum of the intensity of the $G_1$ and $G_2$ features show an almost identical magnitude and spatial variation of the Raman intensity (Figure S6), which suggests an equal enhancement of the Raman intensity in the relevant areas and an equivalent charge distribution in both layers. The slight variation of the Raman intensity can be attributed to the enhancement that occurs due to the variation of the local doping [33].

Using the $I(G_2)/I(G_1)$ ratio, we can estimate the level of delamination in the 1-LG and 2-LG areas. The delamination of the reference top layer ($^{13}$C-1-LG), represented by the median value of the relative intensity of $G_2$, $I(G_2)/I(G_1)$ = 0.35, corresponds to the *AW* ~ 20%, as estimated from the $I(G_2)/I(G_1) = f(AW)$ trend published previously [20]. This coincides with our value of *AW* obtained for the same sample by the AFM analysis (*AW* = 19%). We have to point out that relating the $I(G_2)/I(G_1)$ values to the AFM data for other parts of the MGBL sample is not a perfectly rigorous process, since the general correlation curve ($I(G_2)/I(G_1) = f(AW)$) [20] is derived from a system with a different architecture and therefore a different local charge distribution.

Another important point that should be addressed is the experimentally observed broadening of the Raman bands; the results obtained by the statistical analysis of the Raman maps are summarized in Table S1. The intrinsic broadening of the G and 2D features of 1-LG occurs at about 5 cm$^{-1}$ and 17 cm$^{-1}$, respectively [29]. However, the measured values often exceed the theoretically predicted values by twofold. In the case of the G-mode, the broadening is rationalized by electron–phonon interactions and doping effects [33, 34]. In the case of the 2D mode, the values above 30 cm$^{-1}$ can hardly be attributed only to doping, and effects such as the distribution of the type and magnitude of the strain must be considered [30, 35, 36]; furthermore, the twist angles in the case of 2-LG must be determined [33]. As reported by Neumann et al. [37], and in accordance with our previous work [20], the broadening of the 2D feature can be strongly influenced by the presence of nanoscale wrinkles. Thus, the large values of the 2D broadening ($\delta$), represented by the half-width at half maxima (HWHM intensity; $\delta = 2$HWHM) in our study can be associated with the topographic corrugations of the 1-LG at the nanoscale.

We observed that while the HWHM values obtained from the statistical analysis of the maps of the 2D$_1$ mode are sufficiently homogeneous within the resolution of the laser spot, the data for the 2D$_2$ feature related to the delaminated fraction of the 1-LG are very different for the top and bottom layers. The median values of the FWHM of the 2D$_1$ and 2D$_2$ modes for the $^{13}$C-1-LG (top) and $^{12}$C-1-LG (bottom) layers are 13.5 cm$^{-1}$ and 16.8 cm$^{-1}$ for the top layer and 15.3 cm$^{-1}$ and 18.5 cm$^{-1}$ for the bottom layer, respectively. The bottom $^{12}$C-1-LG layer shows a similar homogeneity for the 2D$_1$ fractions, but the top $^{13}$C-1-LG layer exhibits large variations of the HWHM with a median value corresponding to a value up to 3.5× larger than the broadening that was theoretically predicted [38]. Therefore, we can conclude that the top $^{13}$C-1-LG layer of the MGBL nanostructure shows a much higher level of topographic corrugation at the nanometer scale than the bottom 1-LG shows.

*3.3 Correlation analysis*

The level of doping ($n$) and strain ($\varepsilon$), which reflects the topography of the MGBL structure, was finally evaluated for the principal parts of the sample using the G-2D correlation analysis introduced by Lee et al. [39]. For clarity, the data of the $^{13}$C-1-LG were recalculated relative to the Raman shifts of the graphene with a natural abundance of the $^{12}$C and $^{13}$C isotopes according to equation (1).

The correlation diagram can be decomposed into series of *iso*-strain (doping) and *iso*-doping (strain) lines; the lines for the neutral and unstrained 1-LG intersect at point P$_0$ (1582, 2637) cm$^{-1}$ and separate the correlation diagram into regions of allowed and physically irrelevant values. The slope of the biaxial strain line was reported in the range of 2.25–2.8 [40–42], and the slope of the doping line is about 0.7 [36]. Each point in the correlation diagram can thus be described as a linear combination of the unit vectors corresponding to the strain, $\mathbf{e}_\varepsilon$, and doping, $\mathbf{e}_n$, with the origin of P$_0$.

The Raman shifts of G ($\omega_G$) and 2D ($\omega_{2D}$) are sensitive to both the doping and strain, but with very different fractional variations, ($\Delta\omega_{2D}/\Delta\omega_G$). For the biaxial strain approximation, we can assess the magnitude and sign of the strain within the doping line considering the $\Delta\omega_{2D}$ about -144 cm$^{-1}$ for a 1% strain [39]. The doping can be estimated from the $\Delta\omega_G$ value within the strain line using the formula proposed by Das et al. [43] assuming the hole doping to be due to the SiO$_2$ underlayer [44]. For more details on the correlation analysis, please see the Supplementary information file, Figure S8.

Figure 6 represents the correlation diagrams of the Raman shifts of the 2D and G components for all three representative areas of the MGBL structure. The two sets of the G$_1$; 2D$_1$ and G$_2$; 2D$_2$ correlation pairs are plotted with gray and black colors, respectively. In addition to the border lines of zero doping and strain with slopes of 2.45 and 0.7, respectively, with intersection at P$_0$ (solid red—strain [*iso*-doping]; and blue—doping [*iso*-strain] lines), the doping and strain lines crossing each other in the proximity of the G$_1$; 2D$_1$ and/or G$_2$; 2D$_2$ clouds are also shown (dashed lines).

For the reference parts of the $^{12}$C-1-LG and $^{13}$C-1-LG layers, 1-LG in contact with the substrate is represented by the G$_1$;2D$_1$ pairs located along a strain line of moderate compression ($\varepsilon \sim$ -0.05%). The median value intersects with the doping line corresponding to $n \sim 9\times10^{12}$ cm$^{-2}$ and $1\times10^{13}$ cm$^{-2}$ for the $^{12}$C-1-LG (Figure 6b) and $^{13}$C-1-LG (Figure 6a) layers, respectively. The delaminated fraction of the 1-LG represented by the G$_2$;2D$_2$ pairs shows an almost identical fingerprint with a bimodal distribution of the G$_2$;2D$_2$ pairs, which is only divided by the strain line corresponding to a moderate tension ($\varepsilon \sim$ 0.10%). Thus, two types of landscapes are expected to form, as was observed through AFM.

However, the approximation of the doping level is just slightly lower than that which was identified for the G$_1$;2D$_1$ fraction. The G$_2$;2D$_2$ correlation pairs are located between the doping lines for $n \sim 5\times10^{12}$ cm$^{-2}$ and $1.1\times10^{13}$ cm$^{-2}$ for both the $^{12}$C-1-LG and the $^{13}$C-1-LG layers, respectively.

The scenario is very different for the 2-LG area. Here, the data points of the $G_1;2D_1$ pairs tend to extend along a strain line with $\varepsilon \sim -0.1\%$ and accumulate in the vicinity of the doping line of $n \sim 2\times10^{12}$ cm$^{-2}$ (Figure 6c) and $4\times10^{12}$ cm$^{-2}$ (Figure 6d) for the $^{13}$C-1-LG and $^{12}$C-1-LG layers, respectively. In contrast, the $G_2;2D_2$ pairs are located near the neutral point, and the spot is even more localized in the case of the top layer ($^{13}$C-1-LG) for the $G_2;2D_2$ fraction. This suggests that the minority fraction of the graphene in the top layer of the MGBL structure behaves almost as it does in the pristine graphene.

*3.4 Discussion*

In the context of the general 3D ruga diagram [45], the equilibrium ruga phase is determined by the ratio of the moduli of the 1-LG and the substrate, as well as by the mutual adhesion energy and mismatch strain on the interface of the 1-LG and the substrate. In our case, the topography is expected to be driven mainly by the adhesion to the underlayer features due to most of the self-contacts (1-LG to 1-LG) being in the MGBL structure. The underlayer thus includes the adhered and delaminated parts of the underlying 1-LG and the NPs as local sources of corrugation. The presence of the corrugated 1-LG with some additional local modulation by nanometer-sized objects (NPs) then leads to the formation of a 1-LG surface with different charge and strain management issues than those that were observed in previous studies [20, 46]. This statement is corroborated by the fact that the reference areas show a significant amount of topography. Consequently, the distribution of the doping and strain is very similar to that observed in an analogous system composed of 1-LG transferred over NPs on a SiO$_2$/Si substrate [20, 22]. However, clear evidence of delaminates can be identified in all the inspected areas.

Let us now discuss the effects of strain in the top layer, which was found to be topographically corrugated by the NPs that were introduced underneath. The Raman correlation analysis enabled us to disentangle the strain (and doping) contributions of the adhered and corrugated fractions in the top layer. However, the Raman spectra in the VHS regions were strongly renormalized. As the topography of the MGBL structure was found to be homogeneous within the inspected area of about 20×20 mm$^2$, the data recorded outside of the VHS regions can be used to inspect strain-induced phenomena in the corrugated areas.

First, we estimated effective magnetic length, $l_s$, associated with the strain in the pyramidal objects using the formula devised by Low and Guinea [16]:

$$l_s = \left(\frac{a_0 L}{\beta u}\right)^{1/2} ; \quad \beta = \frac{\partial \log t}{\partial \log a} \ (a = a_0), \qquad (2)$$

where $a_0 = 1.4$ Å, $L$ is the dimension of the strain variation $\sim 10$ nm, $\beta \sim 2 - 3$, $\tau \sim 3$ eV is the nearest neighbor coupling energy, and $u$ is the average strain. Using the values of $u$ determined in the delaminated areas ($\varepsilon = 0.1\%$–0.15%), $\beta \sim 2.5$, we obtain $l_s$ in the range of ~1.93–2.36 nm. The pseudomagnetic field strength, $B_s$, can be obtained as:

$$B_s = \frac{c\beta}{a_0 R}, \qquad (3)$$

where $R$ is the bending radius of the top graphene layer, which does not exceed the value corresponding to the NP radius (~ 5 nm), and $c$ is a dimensionless constant in the order of unity [16]. Using equation (3), we obtain an estimate of $B_s$ in the order of units to tens of Tesla (i.e., $B_s = 5.4$ T for $R = 4$ nm, $\beta = 2.5$ and $c = 1$; for a full plot of $B_s = f(R, \beta)$, see Figure S9). Applying additional real magnetic field in the order of few Tesla, space, and energy separation of the non-equivalent valley states can be achieved, which leads to the stabilization of bulk valley polarization [16].

As reported recently by Kun et al., the PMF fluctuations can also induce a phase shift that is suffered by Dirac charge carriers, giving rise to complete intravalley backscattering of Dirac fermions [17]. Consequently, the scattering on the defect-induced local strain fields must be taken into account. Inspecting our data, the integral intensity ratio of the D'- and D-modes appears to be rather homogeneous within the mapped areas of both the 1-LG and 2-LG regions, reaching typical levels of 0.1–0.5 (see Figure S7), in agreement with previous works (i.e., [47]). We did not observe a dramatic increase by a factor of a few hundred, as reported in [17]. As the laser spot is much larger than the typical dimension of our corrugations, the contribution of the defects from the flat and delaminated parts is most likely merged in a single peak. The peak profile of D and D' is rather symmetrical; therefore, a reliable decomposition into subbands is unrealistic. Thus, we conclude that we do not expect intravalley backscattering of Dirac fermions in our structures. This conclusion is supported by the fact that there are obviously no hidden folds present in our MGBL.

Another interesting aspect of the MGBL comes into play when we consider another magnetic entity within the system. The NPs are not merely passive couplers of the 1-LG layers in the MGBL, but they are also magnetically active. Since the NP size is lower than the critical value of a standalone ferromagnetic domain, the so-called single-domain state onset appears due to the prevalence of the domain wall energy relative to the magnetostatic energy. Thus, the stability of a single-domain state is driven by a purely magnetostatic effect. The critical diameter of a ferromagnetic particle for entering the single-domain regime is typically in the order of units of tens of nanometers for common ferromagnets and ferrimagnets. Consequently, a single NP is represented by a giant classical magnetic moment (sometimes termed as a *superspin*). This mimics the behavior of a giant paramagnetic moment when temperatures exceeding the energy barrier are indicated by the term $K_{eff}V$, where $K_{eff}$ is the effective anisotropy constant and $V$ is the NP volume. In contrast to a single-electron or single-molecule spin, the superspin is merely a classical entity with an intrinsic relaxation time fluctuating on the scale of $10^{-11}$–$10^{-9}$ s. The NP superspin acts as a source of a local time-dependent dipolar field and can be easily stimulated using external static or dynamic electromagnetic fields and can be blocked at temperatures below $\sim K_{eff}V/k_B$ [28].

## 4. Conclusions

In conclusion, we report the successful preparation of a MGBL consisting of two isotopically labeled graphene monolayers, in which the top layer was shaped over uniform NPs with a mean diameter of ~10 nm deposited on the bottom layer. Careful analysis of the MGBL topography suggested the coexistence of two dominant landscapes—flat areas and pyramidal objects located at the position of the NPs. Thanks to the Raman spectroscopy mapping and advanced analysis of the large set of experimental data, we were able to detect the specific changes of the strain and the doping in both the top and bottom layers of the MGBL system as well as in the reference monolayer areas. The strong enhancement of the G-mode detected in specific areas of the MGBL pointed to resonance with the twist induced VHSs. However, the sculpturing of the top layer over the NPs resulted in a large strain locally acting on the graphene, suggesting the formation of pseudo-Landau levels in the top layer corresponding to magnetic pseudofields of ~10 T. We unambiguously demonstrated that the VHSs are robust against local graphene corrugations with spatial distribution in the order of ~10 nanometers and that they can placidly coexist with other strains and/or doping-induced features. The MGBL structure also reveals another coexistence of non-trivial phenomena: a classical giant spin with an intrinsic relaxation time in the order of $10^{-9}$ s and complex gauge fields at the same position. Modification of the MGBL by using magnetic clusters with the spin Hamiltonian specified by the Berry curvature Chern number [48] thus offers an analogous concept at the purely quantum limit. Regarding future perspectives of the MGBL with implanted nanomagnets, further advancements in the physics of the *twist and shape* of 2D lattices can be attained, for example, by spatial confinement and modulation of unconventional superconductivity via time-dependent dipolar fields, giving rise to novel concepts for antimagnets [49] at the nanoscale.

## Acknowledgements

This work was supported by the European Research Council (ERC Starting Grant: 716265) and the Czech Science Foundation (18-20357S). M.S.D. acknowledges support from the U.S. National Science Foundation, DMR-6931929.

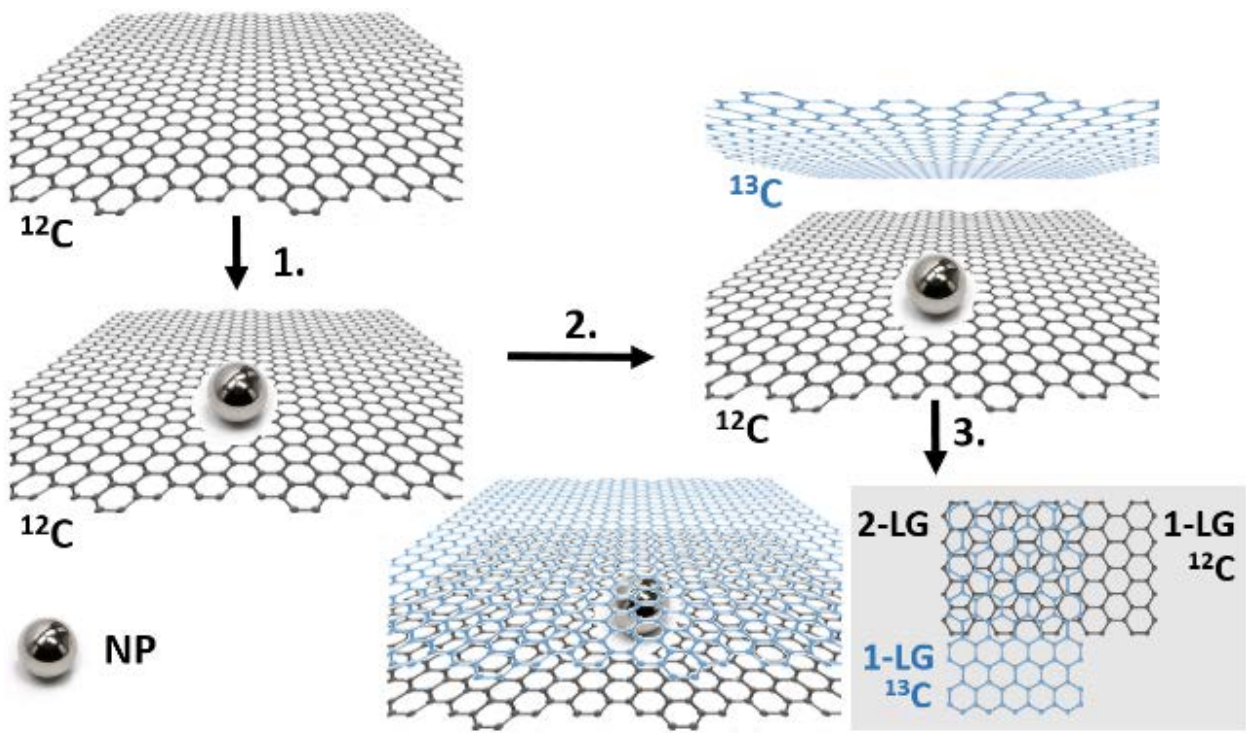

*Figure 1. Schematic representation of the preparation procedure of the MGBL. The process starts with a transfer of the $^{12}C$-1-LG onto a substrate, followed by the deposition of the NPs (1.). The $^{13}C$-1-LG layer is transferred next on top of the decorated $^{12}C$-1-LG layer (2.), and MGBL is created with locally strained $^{13}C$-1-LG due to the presence of NPs (3.).*

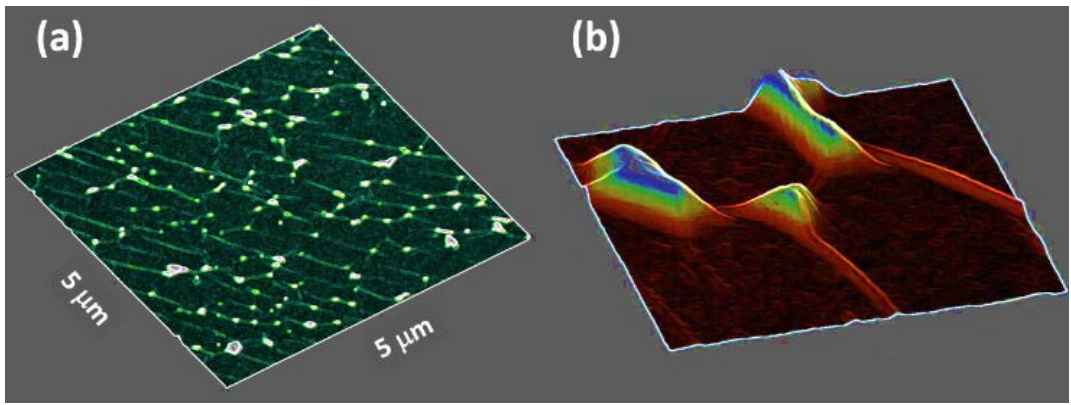

*Figure 2. A large-scale AFM image of the MGBL structure (a). A detail of the features in the vicinity of the NP with vertical and lateral dimensions of ~10 nm (b).*

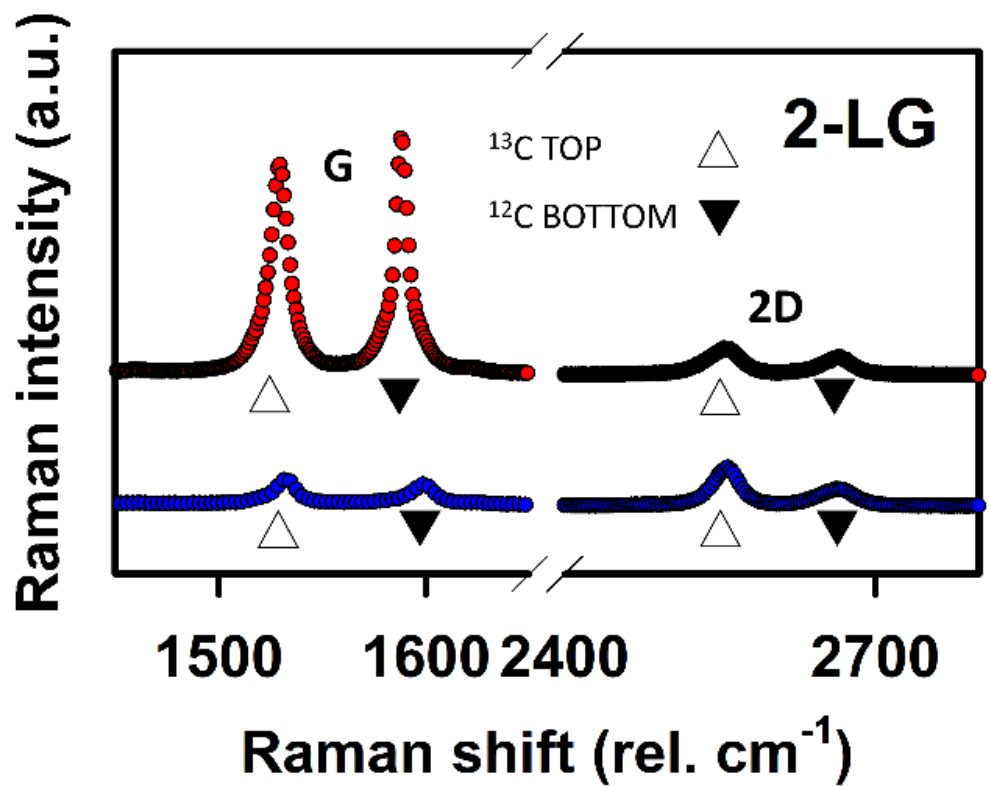

*Figure 3. Raman spectra recorded in the 2-LG region corresponding to the randomly oriented layers (blue points) and the layers oriented in angle ~ 10º (red points), respectively.*

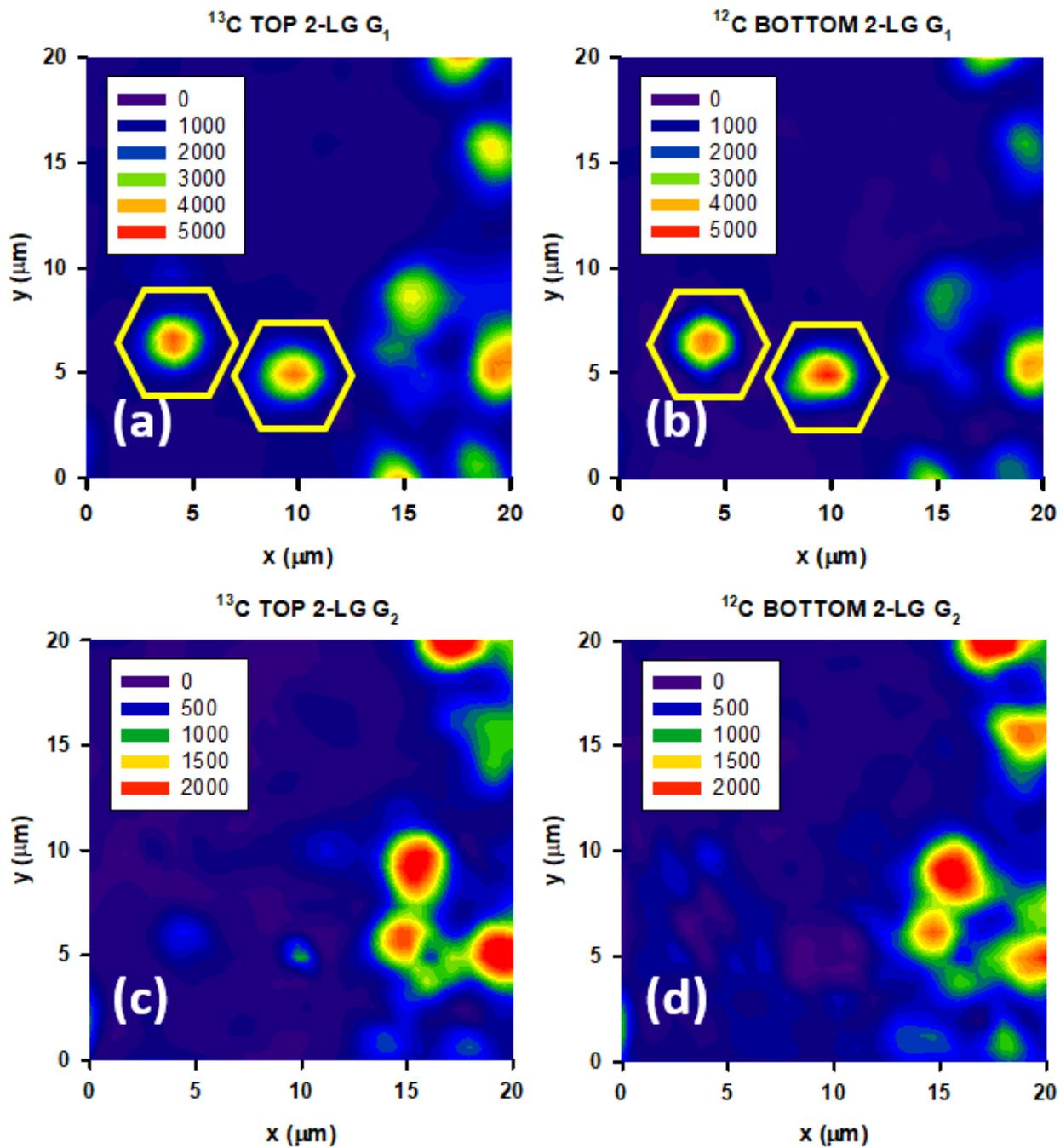

*Figure 4. Large area maps of the Raman intensities of the G-mode for the top (a) and bottom (b) layers, respectively. The VHS regions with a size of several μm² are evidenced by the strong enhancement of the Raman intensity.*

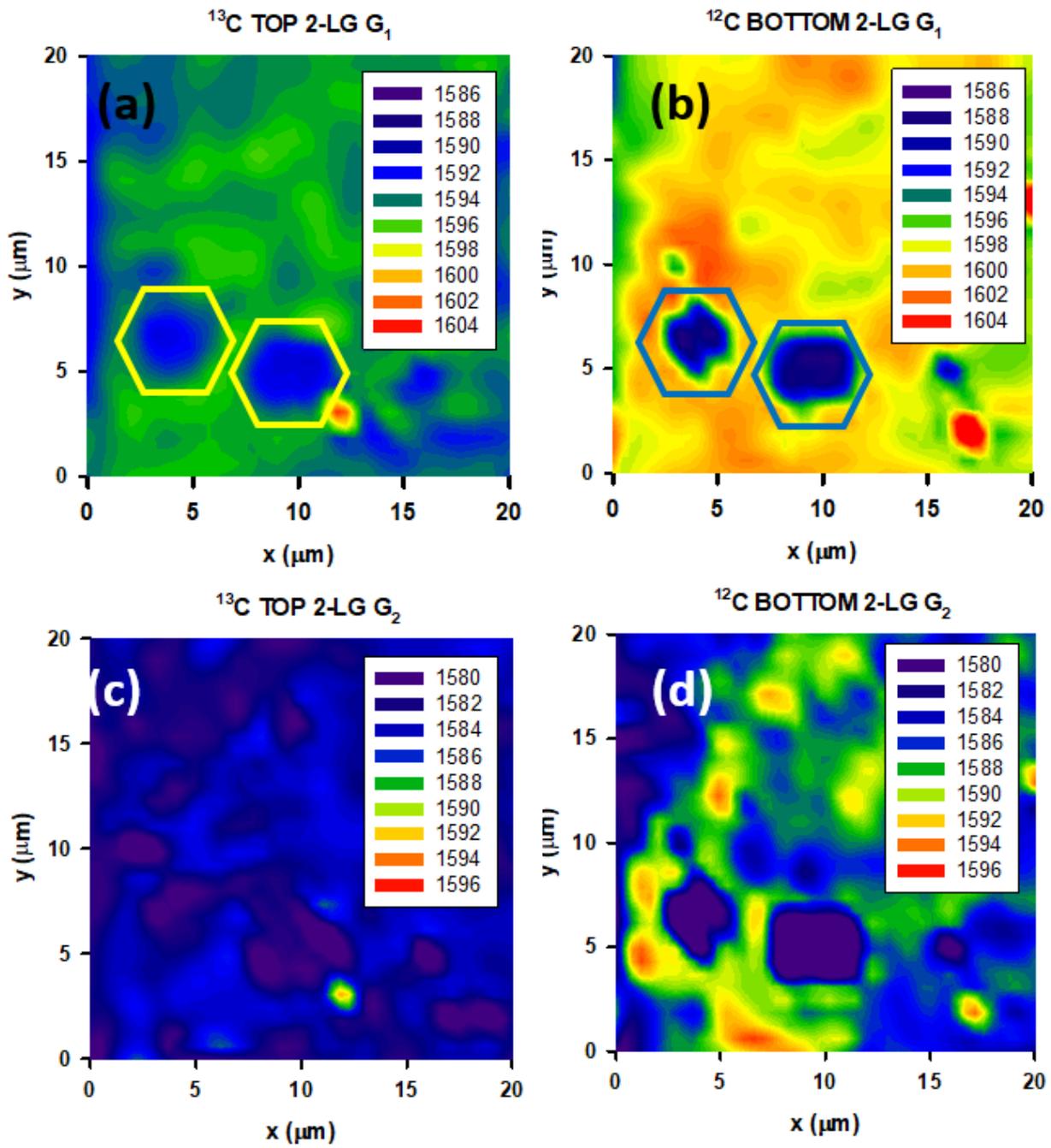

*Figure 5. Large area maps of the Raman shifts of the G-mode for the top (a) and bottom (b) layers, respectively. The VHS regions with a size of several μm$^2$ are evidenced by the areas with low Raman shift values.*

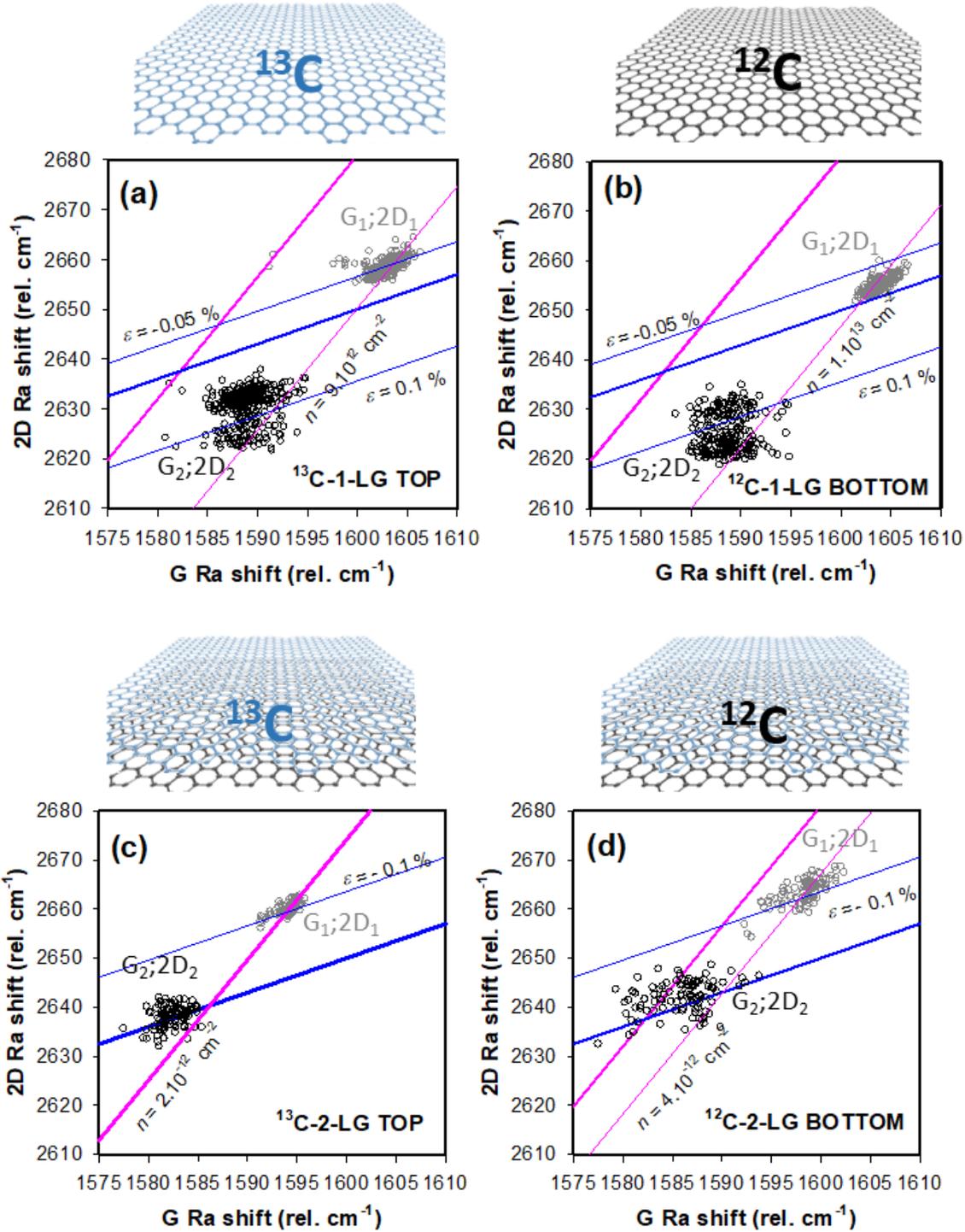

*Figure 6. G-2D correlation plots obtained from the Raman maps recorded on the 1-LG and 2-LG areas out of the Van Hove regions: (a) $^{13}$C-1-LG, (b) $^{13}$C-2-LG, (c) $^{12}$C-2-LG, and (d) $^{12}$C-2LG. Solid gray lines represent the border lines of zero doping with a slope of 2.45 and zero strain with a slope of 0.7, respectively. The two border lines intersects at the (1582, 2637 cm$^{-1}$) point, corresponding to the pristine graphene monolayer. The sample intrinsic doping (iso-strain) and biaxial strain (iso-doping) lines are represented by thin gray lines. The two groups of points, $G_1$; $2D_1$ and $G_2$; $2D_2$, correspond to the adhered and delaminated fractions, respectively. For more detailed definition on the phase space used in the correlation diagrams, please see the Supplementary Information file, Figure S8.*